# Tripartite Phonon-Magnon-Plasmon Coupling, Parametric Amplification, and Formation of a Phonon-Magnon-Plasmon Polariton in a Two-Dimensional Periodic Array of Magnetostrictive/Plasmonic Bilayered Nanodots

Sreya Pal,[1] Pratap Kumar Pal,[1] Raisa Fabiha,[2] Supriyo Bandyopadhyay,[2,*] and Anjan Barman[1,†]

[1]*Department of Condensed Matter and Materials Physics,*
*S. N. Bose National Centre for Basic Sciences, Block–JD, Sector–III, Salt Lake, Kolkata, 700106, India*
[2]*Department of Electrical and Computer Engineering,*
*Virginia Commonwealth University, Richmond, VA 23284, USA*

Coupling between spin waves (SWs) and other types of waves in nanostructured magnetic media has garnered increased attention in recent years because of the rich physics and the potential to produce disruptive technologies. Among this family of intriguing phenomena, we recently reported a new one: coupling between SWs and hybridized phonon-plasmon waves, resulting in tripartite coupling of magnons, phonons, and plasmons. Here, this acousto-plasmo-magnonic phenomenon is studied in a two-dimensional periodic array of bilayered [Co/Al] nanodots on a silicon substrate, where the Co is a magnetostrictive constituent responsive to magneto-elastic coupling and the Al acts as a source of surface plasmons. Time-resolved magneto-optical-Kerr-effect microscopy revealed parametric amplification and strong coupling between two spin wave modes mediated by a hybrid phonon-plasmon wave. The strong coupling forms a new quasi-particle: the phonon-plasmon-magnon polariton.

## I. INTRODUCTION

Plasmonics [1, 2] and magneto-plasmonics [3, 4] have been busy fields of study, spawning novel technologies such as magneto-optical nano-antennas [5, 6] and circuits [7]. In magneto-plasmonics, one usually examines the influence of magnetic fields on plasmons, while ignoring the reciprocal effect where surface plasmons, potentially hybridized with entities like phonons [8–10], impact the dynamic magnetic properties of magnetic media, specifically spin waves. Here, we report a study of this latter effect which we recently showed could reveal intriguing features such as the formation of acousto-plasmo-magnonic frequency combs [11]. This field is an extension of the well-explored and popular field of bipartite magnetoelastic coupling [12–21] where pure phonon modes interact with magnon modes (magnon-phonon coupling) in magnetostrictive nanomagnets to produce such entities as magnon-polarons [20, 21]. The next frontier is obviously tripartite coupling where three different entities — phonons, magnons and plasmons – couple to reveal new physics and portend new applications.

Conventional wisdom will hold that phonon-magnon-plasmon coupling would be an elusive phenomenon since plasma frequencies far exceed those of spin waves. Hence, any coupling between them will encounter a nearly insurmountable barrier because of the resulting large phase mismatch. Surface plasmons however can have much lower frequencies than bulk plasmons and hence they can couple with acoustic phonons in appropriate media to form hybrid phonon-plasmon modes [8–10] whose frequencies are close to those of spin waves in ferromagnetic media. These hybrid modes [phonons + plasmons] can couple strongly with spin wave modes [magnons] to produce tripartite coupling between phonons, magnons and plasmons, giving rise to acousto-plasmo-magnonics.

In this paper, we have employed time-resolved magneto-optical Kerr effect (TR-MOKE) microscopy to showcase the intriguing coupling between spin waves and hybridized phonon-plasmon waves. Our study employs a two-dimensional periodic array of ∼100 nm [Co/Al] heterostructured nanodots deposited on a Si substrate. Acousto-plasmo-magnonic modes are excited in this system when it is exposed to ultrashort (∼100 fs) laser pulses. The periodic heating and cooling caused by the laser pulses give rise to acoustic waves (phonons) in the Si substrate owing to periodic stresses caused by the unequal thermal expansion/contraction coefficients of the Si substrate and the nanodots on top. These phonons mix with the plasmons emitted by the Al layer to form the hybrid phonon-plasmon modes, which then couple with the spin wave modes generated in the nanomagnets, with or without an external magnetic field, to form the acousto-plasmo-magnonic modes whose dispersion relations may show no dependence on any magnetic field.

## II. EXPERIMENTS

### A. Sample Fabrication

The samples are fabricated with e-beam lithography and sequential e-beam evaporation of different metals. An e-beam resist is spun onto a Si substrate and patterned with e-beam in a Raith system to open windows for the nanodots, followed by sequential evaporation of Ta (10 nm: for adhesion), Al (10 nm), Co (6 nm) and Au (2 nm), followed by lift-off. The Au layer is needed to prevent oxidation of Co. We fabricate two sets of samples on Si substrates: (i) a two-dimensional array of elliptical Co nanomagnets [NMs] and (ii) an identical array of [Co/Al] bilayered nanomagnets of the same shape and dimensions. The former set does not have plasmons (because no plasmonic material is present), but the latter set does.

* sbandy@vcu.edu
† abarman@bose.res.in

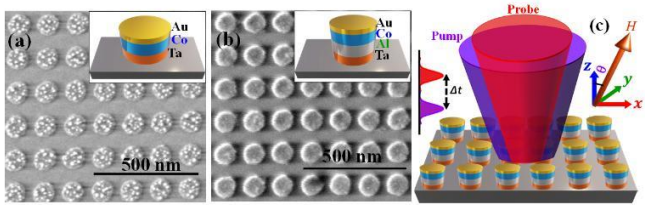

Figure 1. (a) Scanning electron microscope (SEM) images of the Co nanomagnet (NM) arrays on Si substrate. (b) SEM images of the identical [Co/Al] nanodot arrays. (c) Schematic of the TR-MOKE measurement geometry.

### B. Measurement Technique

The investigation of ultrafast magnetization dynamics of the samples is conducted by employing a custom built TR-MOKE microscope [22] based on a two-color collinear pump–probe technique under ambient conditions. A schematic representation of this experimental setup is shown in Fig. 1 (c).

Fig. 1 (a) and Fig. 1 (b) show the scanning electron microscope (SEM) images of systems (i) and (ii), respectively. The Co NMs are shaped like elliptical disks with major axis dimension ∼105 nm, minor axis dimension ∼100 nm, and thickness ∼6 nm. Along the major axes (x-axis), the edge-to-edge separation between the NMs is ∼45 nm and along the direction of the minor axes, it is ∼50 nm. The lateral dimensions of the nanomagnets and the edge-to-edge separation between them exhibit a maximum deviation of ±5% and ±8%, respectively.

The fundamental laser beam (wavelength = 800 nm, fluence = 2 mJ cm$^{-2}$, pulse width = 80 fs) from a mod-elocked Ti-sapphire laser (Tsunami, Spectra-Physics) is used to probe the polar Kerr rotation (hereafter referred to as the probe beam) and its frequency-doubled counterpart (wavelength = 400 nm, fluence = 16 mJ cm$^{-2}$, pulse width = 100 fs) is used to excite the magnetization dynamics of the samples. The Kerr signal is measured using an optical bridge detector (OBD) as a function of the time delay between the pump and probe beams. Achieving spatial overlap of these two beams is critical. The slightly defocused pump beam (of diameter ∼1 µm) at the focal plane of the tightly focused probe beam (of diameter ∼800 nm), is made overlapping on the sample plane using a microscope objective of numerical aperture 0.65. The OBD simultaneously measures both the reflectivity and Kerr rotation and separates these intertwined signals with the help of two lock-in amplifiers in a phase-sensitive manner to attain high sensitivity. To maintain temporal synchrony, the pump beam is subjected to periodic modulation at a frequency of about 2 kHz by a mechanical chopper. This modulated pump beam serves as a reference frequency and is conveyed to the lock-in amplifiers, anchoring the phase relationships within the system. Further, the probe beam is systematically positioned at the precise center of the pump beam using an x–y–z piezoelectric scanning controller, guided by a feedback loop and a white-light illumination system, enhancing the fidelity of the experiment. A static magnetic field with varying magnitude is applied with a small tilt (of ∼ 10◦ - 15◦) from the sample plane to introduce the necessary out-of-plane demagnetizing field along the direction of the pump beam, which helps to induce precessional magnetization dynamics (spin waves) in the sample during pumping. The experimental time window of 2 ns was found to be sufficient to resolve the spin wave (SW) peaks with a temporal resolution of 10 ps from the fast Fourier transform (FFT) of the bi-exponential background subtracted time-resolved traces.

### C. Micromagnetic Simulations

We compared experimental results of the samples without Al (no plasmons present) with theoretical results obtained from micromagnetic simulations using OOMMF (Object Oriented Micromagnetic Framework software) [23], employing a 4× 4 array of nanomagnets whose specifications are based on what we observed in the SEM images. The discretized array featured rectangular parallelepiped-shaped cells ($3 \times 3 \times 6$ nm$^3$) with a 2D periodic boundary condition. The unit cell length was kept below the exchange length of Co (∼4.93 nm), to allow the incorporation of exchange interactions. Simulation parameters used included saturation magnetization, $M_s$ = 1400 emu cm$^{-3}$, anisotropy constant, $K = 0$, $\gamma = 17.6$ MHz Oe$^{-1}$, the exchange stiffness constant, $A_{ex} = 3.0 \times 10^{-6}$ erg cm$^{-1}$ [24], and Gilbert damping coefficient, $\alpha$ = 0.008 [25].

To simulate magnetization dynamics, we first obtained the static magnetic state after system relaxation to equilibrium at a specific bias-field. Optical excitation was emulated with a square pulsed magnetic field (10 ps rise and fall time, 200 ps width, 20 Oe peak amplitude) perpendicular to the sample plane. The FFT of the simulated time-resolved out-of-plane magnetization ($m_z$) revealed the SW spectra, consistently showing three dominant peaks (M1, M2, and M3) across various bias magnetic fields. These theoretical results showed excellent agreement with the experimental observations. We were not able to simulate the magnetization dynamics in the samples containing Al since no software is available that can incorporate the effect of plasmons in magneto-dynamics. Therefore, we are unable to compare theory with experiments in the Al-containing samples. Our experiments, however, showed stark differences between the plasmonic (Al-present) and non-plasmonic (Al-absent) samples, which can only accrue from coupling with plasmons.

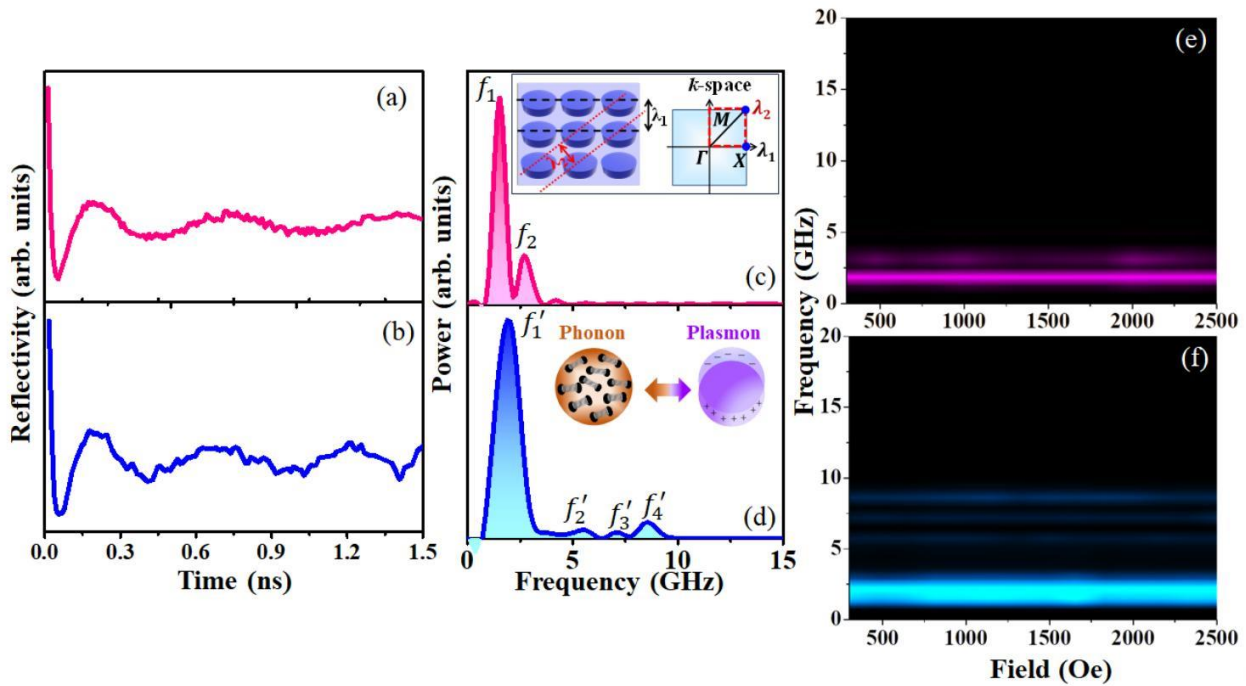

Figure 2. Background-subtracted time-resolved data for the reflectivity of (a) the Co NM arrays and (b) the [Co/Al] NM arrays, on Si substrate as a function of the delay between the pump and the probe, obtained at 16 mJ cm$^{-2}$ pump fluence. c) and d) FFT of the respective oscillations. The two detected frequencies in the non-plasmonic sample correspond to surface acoustic wave modes excited in the Si substrate by the laser heating/cooling, whose wavelengths are commensurate with distinct periodicities in real space and reciprocal space of the magnonic crystal made of the NM arrays [inset of (c)]. These are bending wave modes [pure phonons] excited by the pump beam periodically heating and cooling the substrate. The spectrum of the Co/Al plasmonic sample shows that the primary peak has blue-shifted and its full-width at half maximum has increased considerably, compared to that of the non-plasmonic sample. There are also three satellite peaks. These are hybrid phonon-plasmon modes; the phonons are excited by the pump beam and they mix with plasmons emitted by the Al layer to form the hybrid phonon-plasmon wave. e) and f) Surface plot of the reflectivity spectra of the Co NM arrays on Si substrate and identical [Co/Al] NM arrays, respectively.

### D. Reflectivity Measurements

In Fig. 2 (a) and (b), we show the background subtracted time-resolved data for reflectivity of the non-plasmonic (without aluminium) and plasmonic (with aluminium) samples as a function of the delay between the pump and the probe, obtained at 16 mJ cm$^{-2}$ pump fluence. We will discuss the non-plasmonic results first.

The fast Fourier transformation (FFT) of the oscillations for non-plasmonic samples reveal the dominance of two peaks at 1.5 GHz ($f_1$) and 2.7 GHz ($f_2$) in the spectra. These modes are absent for the case of FFT of time resolved oscillations from bare substrate (without NM).

The origin of the modes in the non-plasmonic sample can be explained by invoking bending waves in the samples due to unequal thermal contraction/expansion of the NMs and substrate when heated by the pump laser pulse. The periodic strain produced by the unequal expansion/contraction spawns the bending waves (BW). For simplicity, we will assume a vertical rod-like geometry for the NMs in our analysis. When the wavelength of the BW is large compared with the thickness of the rod and the oscillations are in a direction perpendicular to the axis of the rod, the frequency ($\omega$) and the wavevector ($k$) obey the dispersion relation [26]

$$\omega = k^2 \sqrt{\frac{Et^2}{12\rho(1-\sigma^2)}} \quad (1)$$

where $E$, $\rho$ and $\sigma$ are the Young's modulus, density and the Poisson's ratio of Co, respectively, and $t$ is the thickness of the Co nanodot (6 nm). The material parameters for Co are: $E = 200$ GPa, $\rho = 8900$ kg m$^{-3}$ and $\sigma = 0.32$ [27]. The dominant peak frequency $f_1$ observed in Fig. 2 (= 1.5 GHz) obeys Equation (1), when the wave vec-tor corresponds to the X point of the 2D Brillouin zone (BZ) $|k_1| = \frac{2\pi}{|l|}$ (l is the NM array periodicity). The secondary peak frequency $f_2$ (= 2.7GHz) corresponds to bending waves with the wave vector along the diagonal of the two-dimensional lattice (M point of the BZ), i.e.

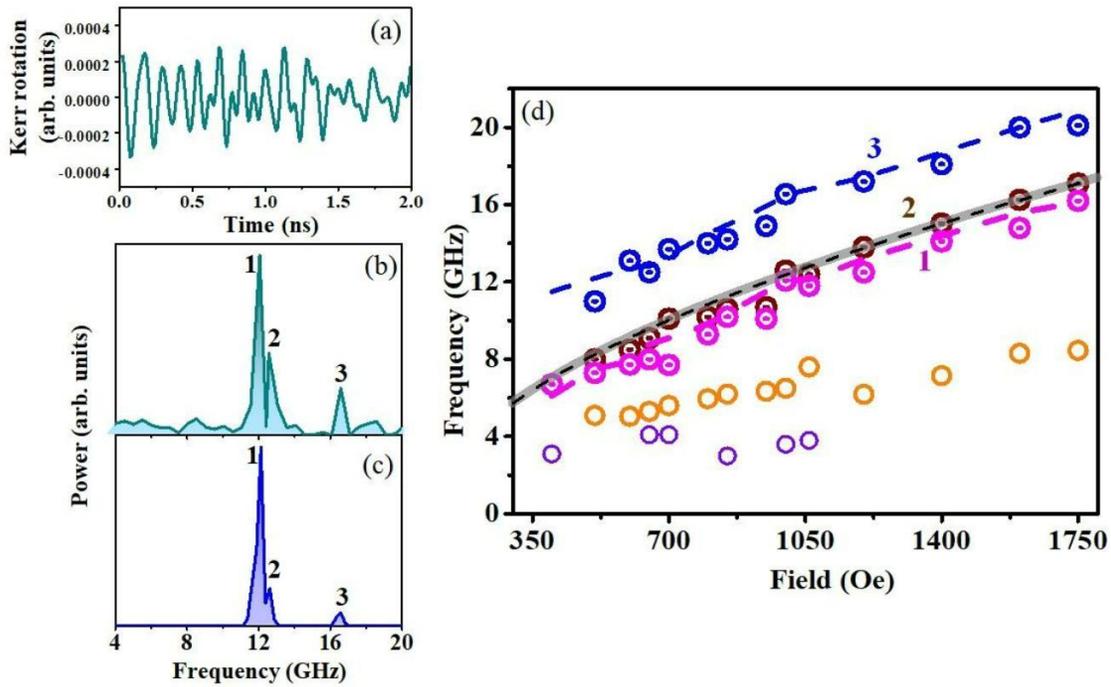

Figure 3. (a) Background subtracted time-resolved Kerr oscillations from Co NM arrays on Si substrate (no Al present) at the pump fluence of 16 mJ cm$^{-2}$, b) FFT of the corresponding oscillations. c) FFT of the simulated temporal evolution of the $m_z$ component (out-of-plane component of magnetization) at a magnetic field of 1 kOe using OOMMF, showing good agreement with the experimentally measured spectra in (b). d) Bias magnetic field dependence of the observed three dominant frequencies in the Kerr oscillations. The circles represent experimental data and the solid line is Kittel fit for mode '2'. The dashed lines are micromagnetic simulation results.

$|\vec{k_2}| = \sqrt{2}\frac{2\pi}{|l|}$. The calculated velocity of propagation of the bending waves for this periodicity of the system is of the order of 1 km s$^{-1}$. For the plasmonic sample, FFT of the oscillations reveals the dominance of four peaks at 2 GHz ($f_1'$), 5.5 GHz ($f_2'$), 7.1 GHz ($f_3'$), and 8.6 GHz ($f_4'$), where the intensities of the last three are much lower than that of the first. The observed frequencies $f_n$ (n= 1,2) [non-plasmonic sample] and ($f_n'$) (n = 1,2,3,4) [plasmonic sample] are insensitive to the presence or absence of any magnetic field at any given fluence and show exceptional stability over the broad range of magnetic field, confirming the non-magnetic origin of these oscillations. The emergence of ($f_n'$) modes is contingent upon the presence of the Al layer, thereby implicating their plasmonic origin.

Apart from the fact that the non-plasmonic sample has two frequency peaks and the plasmonic sample has four, two differences between the spectra in the non-plasmonic and plasmonic samples [Fig. 2(c) and (d)] stand out. The dominant peak in the non-plasmonic sample is blue shifted in the plasmonic sample, and the linewidth has increased. It appears that the hybrid phonon-plasmon mode born out of the mixing of bending waves with plasma waves in the plasmonic sample (which we are observing in Fig. 2(d)) is primarily phononic in nature, i.e. the bending wave dominates, because the frequencies of the dominant modes in the two samples – one purely phononic (BW) and the other plasmo-phononic – are somewhat close to each other. The small blue shift in the plasmonic sample implies that the quasi-particle in the hybrid phonon-plasmon mode is more energetic than the pure phonon in the BW. This is not surprising since the phonon can absorb some energy from the plasmon to cause the blue shift in the plasmonic sample.

On the other hand, the increased linewidth indicates that the hybrid phonon-plasmon mode is more lossy than the pure phonon (BW) mode. Plasmons couple very strongly with electromagnetic waves [28] and hence the hybrid phonon-plasmon wave may radiate some photons (electromagnetic waves) causing more rapid loss of energy. This is a speculation and we intend to verify this in future by measuring any possible electromagnetic wave emission from these samples.

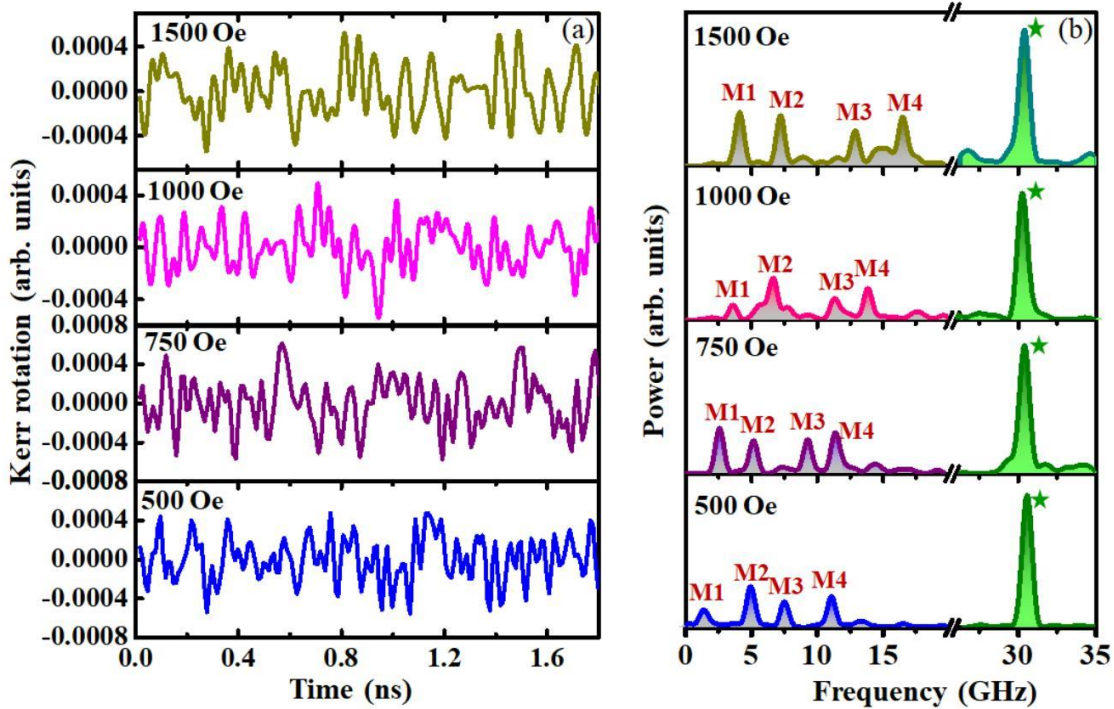

Figure 4. (a) Bias magnetic field dependence of the background subtracted time-resolved Kerr oscillations from [Co/Al] NM arrays at the pump fluence of 16 mJ cm$^{-2}$. b) The FFT of the corresponding Kerr oscillations showing a field independent mode at 30.5 GHz, which is independent of the bias magnetic field.

### E. Time-resolved Kerr oscillation measurements

Fig. 3 shows the experimentally measured background subtracted Kerr oscillation data obtained from the Co nanomagnets on Si (no Al present) at different bias magnetic fields. The observed peaks in the FFT spectra exhibit a systematic upward shift in frequency as the bias magnetic field increases, confirming their magnetic origin. Despite a modest signal-to-noise ratio in the data, three distinct peaks denoted as '1', '2', and '3' have been discerned in the FFT spectra. We could fit the magnetic field dependence of mode '2' with the Kittel formula [29]

$$f = \frac{\gamma}{2\pi}\sqrt{H(H + 4\pi M_{eff})} \quad (2)$$

where the gyromagnetic ratio, $\gamma = g\mu_B/\hbar$, H is the bias magnetic field, $g$ is the Landé g factor ($g=2$), $\mu_B$ is the Bohr magneton, $\mu_0$ is the permittivity of vacuum, and $\hbar$ is the reduced Planck's constant. The fitting results are depicted in Fig. 3(d), demonstrating that mode '2' conforms well to a Kittel-like mode. The effective magnetization ($M_{eff}$) required for the fit was 1250 emu cm$^{-3}$, a value slightly lower than the intrinsic saturation magnetization of Cobalt (Co) at 1400 emu cm$^{-3}$. This discrepancy is attributed to a somewhat non-uniform magnetization profile within each nanomagnet.

In contrast, modes '1' and '3' cannot be fitted with either the Kittel formula or the perpendicular standing spin wave (PSSW) mode relation. Mode '1' is likely an edge mode, while mode '3' is identified as a hybrid magneto-dynamical mode, consistent with previous reports in these systems [27].

In stark contrast to the non-plasmonic sample with no Al present, the sample with Al (where plasmonic coupling is present) exhibits much richer spectra. One easily discernible high frequency magnetic field-independent mode at ~30.5 GHz is always present, in addition to other field-dependent modes. The former has more power than all other modes. This field-independent mode is only present when the Al layer is there, confirming its plasmonic origin. Remarkably, it remains stable regardless of the presence or absence of a magnetic field. Fig. 4(b) illustrates this stable plasmonic mode across four different bias magnetic fields (it is the mode marked with an asterisk).

Fig. 5 (a) reveals the presence of field-dependent modes in the plasmonic Al-containing sample in addition to the high-frequency field-independent mode at 30.5 GHz. The dispersions of modes M4 and M3 are fitted by the blue and red lines with arrowheads. These lines are merely guide to the eye. A distinct frequency gap of magnitude 2g opens up at a magnetic field of 820 Oe. This gap attests to the coupling between M4 and M3, which is obviously mediated by the hybrid phonon-plasmon mode.

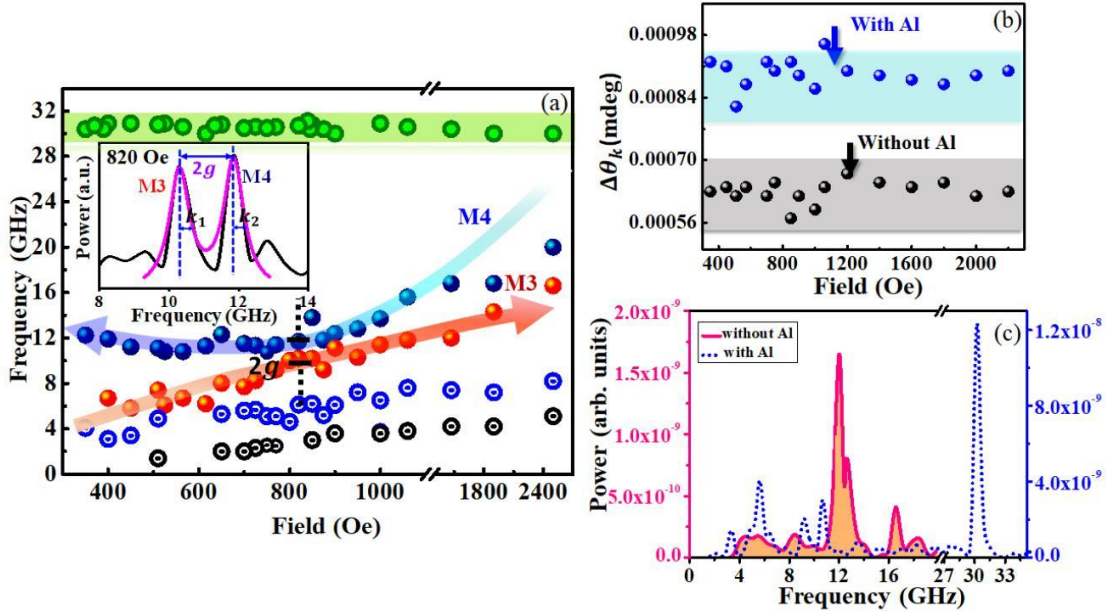

Figure 5. (a) Experimental bias magnetic field dependence of Kerr oscillation modes in [Co/Al] NM arrays [plasmonic samples]. The lines with arrow heads are guide to the eye for the modes M4 and M3. A gap of frequency $2g$ opens up between these two dispersion curves at a magnetic field of 820 Oe, showing the existence of coupling between them at that magnetic field. The inset shows the FFT power spectrum of the observed field-independent and field-dependent modes at the magnetic field of H = 820 Oe (black solid line). Here, $2k_1$ and $2k_2$ are the full-width-at-half-maximum (FWHM) of the Lorentzian fits to the spectra of M4 and M3. b) Average peak-to-peak Kerr rotation amplitude (averaged over all modes) in the plasmonic sample (with Al) and in the non-plasmonic sample (without Al), as a function of the bias magnetic field. c) Amplitude of power in the FFT spectra obtained from Kerr rotation of plasmonic and non-plasmonic samples at magnetic field H = 1 kOe.

The gap $2g$ and the loss rates (half width at half maximum) of the modes M4 and M3, denoted as $k_1$ and $k_2$, are depicted in the inset of Fig. 2(a). To quantify these parameters, we applied a Lorentzian fit to the peaks corresponding to M4 and M3. The extracted values are, $g$ =0.78 GHz, $k_1 = 0.36$ GHz and $k_2 = 0.35$ GHz. The cooperativity factor for coupling is defined as $C = \frac{g^2}{k_1 k_2}$, which, in our case, has a value of 4.8. In accordance with established criteria, strong coupling is delineated by $C > 1$ and $g > k_1, k_2$. Because we satisfy both conditions, we have strong coupling [30].

In a regime of strong coupling, individual modes transform, giving rise to a novel quasi-particle, akin to the magnon-polaron observed in spin wave coupling with pure acoustic waves [20, 21]. Our system exhibits robust tripartite coupling involving phonons, plasmons, and magnons, suggesting the emergence of a distinct phonon-plasmon-magnon polariton – a new quasi-particle. This phenomenon, reminiscent of strongly coupled magnon-plasmon polaritons reported in graphene-2D ferromagnet heterostructures [31], underscores the significance of plasmons in our system.

Notably, the absence of coupling in the sample without aluminium (no plasmons) is evidenced by the lack of anti-crossing (gap opening) in the dispersion curves for field-dependent modes. This unequivocally demonstrates that plasmons introduce qualitative distinctions, as also seen in the reflectivity spectra, not merely quantitative ones, and coupling between spin wave modes occurs in their presence because the hybrid phonon-plasmon wave mediates the coupling.

### F.  Parametric amplification of acousto-plasmo-magnon spin wave modes

Our experiments have unveiled two other intriguing findings: First, the Magneto-Optical Kerr Effect (MOKE) signal, specifically the average peak-to-peak Kerr rotation amplitude across all modes at an arbitrarily picked bias magnetic field of 1 kOe, exhibits a notable 30% increase in the plasmonic sample with Al compared to the magnetic field-dependent modes in the non-plasmonic sample without Al. This enhancement persists across a broad range of magnetic fields. Second, the amplitude of power in the FFT spectra corresponding to the anti-crossed acousto-plasmo-magnon spin wave modes in the Al-containing sample is approximately four times greater than that in the field-dependent modes of the Al-lacking sample when averaged across all modes. Notably, the intensity of the field-independent mode is ten times higher than that of the field-dependent modes, as illustrated in Fig. 5(c).

The significant increase in the MOKE signal, specifi-cally a 30% rise in the average peak-to-peak Kerr rotation amplitude across all modes in the sample with aluminium (Al) compared to the Al-lacking sample, suggests the presence of parametric amplification. In this context, energy is transferred from the hybridized phonon-plasmon mode to the intrinsic spin wave modes in the Al-containing sample, thereby amplifying the amplitude and power of the resultant acousto-plasmo-spin wave mode. This observation serves as confirmation of the robust tripartite plasmon-phonon-magnon coupling, since such power transfer can only happen if there is strong coupling.

### III. CONCLUSION

In conclusion, we report (i) tripartite coupling between phonons, plasmons and magnons in periodic arrays of bilayered nanodots of a magnetostrictive material and a plasmonic material, spawning a high frequency magnetic field independent acousto-plasmo spin wave mode; (ii) parametric amplification of spin wave modes by hybrid phonon-plasmon modes, and (iii) strong coupling between two spin wave modes mediated by the hybrid phonon-plasmon wave, leading to the formation of the magnon-plasmon-phonon polariton. These findings not only contribute to our understanding of fundamental interactions in magnonic and plasmonic systems, but also hold significant implications for practical applications. The observed parametric amplification can be leveraged for spin-wave device applications. Furthermore, these results open avenues for controlling spin dynamics through the interplay of plasmon-phonon-photon dynamics. This approach has the potential to mitigate intrinsic limitations associated with ohmic and other losses, offering exciting prospects for plasmon-driven coherent spintronics or magneto-optical activities.

### ACKNOWLEDGMENTS

A.B. gratefully acknowledges Department of Science and Technology, Govt. of India (grant no. DST/NM/TUE/QM-3/2019-1C-SNB) for financial assistance. A.B. and S.B. acknowledge support from the Indo-US Science and Technology Fund Center grant "Center for Nanomagnetics for Energy-Efficient Computing, Communications, and Data Storage" (IUSSTF/JC-030/2018). The work of R.F. and S.B. was supported by the US National Science Foundation under grant ECCS-2235789. S.P. and P.K.P. acknowledge the Council of Scientific and Industrial Research (CSIR), Govt. of India for respective senior research fellowships.

(Supplemental Material)

# Tripartite Phonon-Magnon-Plasmon Coupling, Parametric Amplification, and Formation of a Phonon-Magnon-Plasmon Polariton in a Two-Dimensional Periodic Array of Magnetostrictive/Plasmonic Bilayered Nanodots

Sreya Pal,[1] Pratap Kumar Pal,[1] Raisa Fabiha,[2]
Supriyo Bandyopadhyay,[2,*] and Anjan Barman[1,†]

[1]*Department of Condensed Matter and Materials Physics,*
*S. N. Bose National Centre for Basic Sciences,*
*Block–JD, Sector–III, Salt Lake, Kolkata, 700106, India*

[2]*Department of Electrical and Computer Engineering,*
*Virginia Commonwealth University, Richmond, VA 23284, USA*

# I. SPIN WAVE MODE ANALYSIS FOR CO NANOMAGNETS (NON-PLASMONIC SAMPLES

To understand the characteristics of three magnetic modes in the samples without Al, including the Kittel-type and other modes, we used an in-house simulator Dotmag[1] to calculate their spatial properties (power and phase). The results, presented in Fig. S1, reveal distinct features. For mode '2', the power profile indicates a uniform distribution of spin wave (SW) power at the nanomagnet's center, resembling a uniform precessional mode or center mode (CM). Despite high power at the edges, the central distribution is significant. In contrast, mode '1' shows power concentrated at the edges, termed the edge mode (EM), with power focused at two vertical edges perpendicular to the bias field. Mode '3' is akin to '2', except the quantization axis rotates about 45° with respect to the major axis of the elliptical nanomagnet. Analyzing phase profiles, we observe spins precessing alternately in opposite phases, creating azimuthal contrast with an azimuthal mode quantization number of 2. Notably, power distributions for these modes (both at the center and vertical edges of the nanomagnet) remain remarkably stable, showing minimal change with variations in external bias field strength.

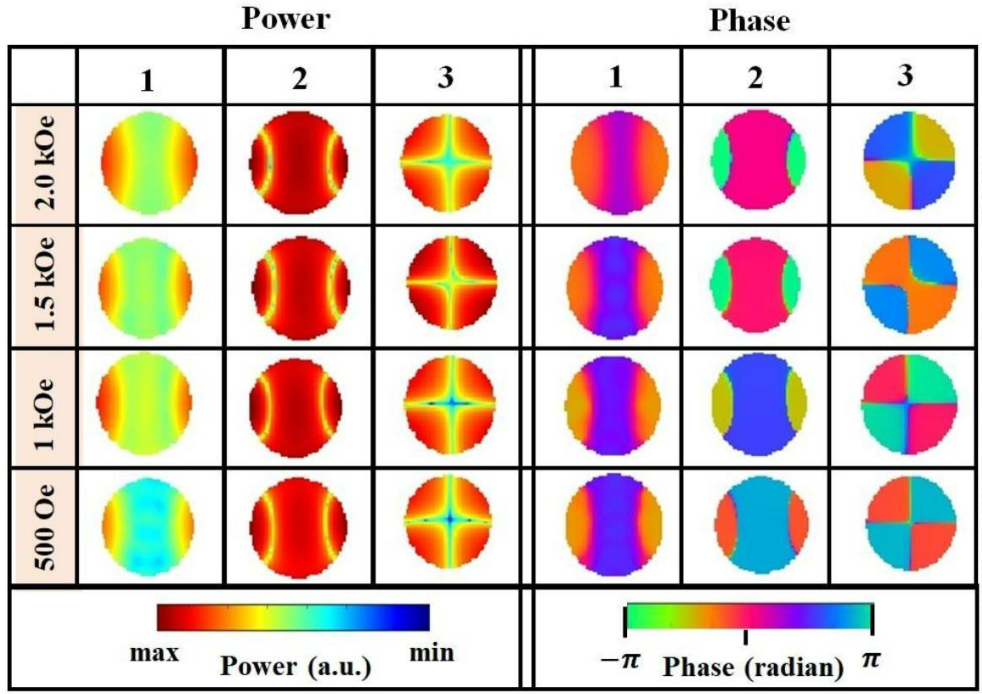

FIG. S1. Simulated power and phase profiles of different spin wave modes (Mode 1, mode 2, mode 3) for Co nanomagnets without Al at four bias magnetic fields. Colour maps for power and phase distributions are shown below the images.

---